\documentclass[12pt]{article}

\usepackage{amsmath}
\usepackage{amssymb}
\usepackage{euscript}

\usepackage{epsfig}

\usepackage{cite}

\usepackage{alltt}

\def\Order#1{${\cal O}(#1$)}

\begin{document}                     

\allowdisplaybreaks

\begin{titlepage}

\begin{flushright}
{\bf  CERN-TH/2001-228\\
      MPI-PhT-2001-28\\
      UTHEP-01-0801
}
\end{flushright}

\vspace{1mm}
\begin{center}
{\LARGE
On Theoretical Uncertainties of the \\
\vspace{1mm}
$W$ Boson Mass Measurement at LEP2$^{\star}$
}
\end{center}

\vspace{1mm}
\begin{center}
{\bf S. Jadach$^{a,b}$, W. P\l{}aczek$^{c,b}$, M. Skrzypek$^{a,b}$ 
        B.F.L. Ward$^{d,e,f}$} {\em and} {\bf Z. W\c{a}s$^{a,b}$}

\vspace{1mm}
{\em
$^a$Institute of Nuclear Physics,
  ul. Kawiory 26a, 30-055 Cracow, Poland,\\
$^b$CERN, TH Division, CH-1211 Geneva 23, Switzerland,\\
$^c$Institute of Computer Science, Jagellonian University,\\
   ul. Nawojki 11, 30-072 Cracow, Poland,\\
$^d$ Max-Planck-Institut f\"ur Physik, 80805 Munich, Germany\\
$^e$Department of Physics and Astronomy,\\
  The University of Tennessee, Knoxville, TN 37996-1200, USA,\\
$^f$SLAC, Stanford University, Stanford, CA 94309, USA.
}
\end{center}

\vspace{5mm}
\begin{abstract}
We discuss theoretical uncertainties of the measurement of the 
$W$ boson mass
at LEP2 energies, reconstructed with the help of the tandem of the
Monte Carlo event generators KoralW and YFSWW3.
Exploiting numerical results obtained with these programs,
and the existing knowledge in the literature,
we estimate that the theoretical uncertainty of the $W$ mass
due to electroweak corrections,
as reconstructed at LEP2 with the help of these programs,
is $\sim 5\,$MeV. 
Since we use certain idealized event selections and 
a simple $M_W$-fitting procedure, our numerical exercises can be
(should be) repeated for the actual ``$M_W$ extraction methods'' 
of the LEP2 measurements, using KoralW and YFSWW3 or other Monte Carlo
programs.
\end{abstract}

\vspace{5mm}
\begin{center}
{\it To be submitted to Physics Letters B}
\end{center}

\vspace{12mm}
\footnoterule
\noindent
{\footnotesize
\begin{itemize}
\item[${}^{\star}$]
  Work partly supported by 
  the Polish Government grants KBN 5P03B12420 and KBN 5P03B09320,
  the European Commission 5-th framework contract HPRN-CT-2000-00149,
  the US DoE Contracts DE-FG05-91ER40627 and DE-AC03-76ER00515.
\end{itemize}
}

\vspace{1mm}
\begin{flushleft}
{\bf CERN-TH/2001-228\\
      MPI-PhT-2001-28\\
      UTHEP-01-0801
\\  August 2001
}
\end{flushleft}

\end{titlepage}

In this work we would like to present our estimate of the theoretical 
uncertainties (TUs) related to electroweak corrections in the measurement 
of the mass of the $W$ heavy boson in the LEP2 experiments.
The estimate will be based on new numerical results of our own and on the best
results available in the literature.
One important reason for writing this paper is that the
discussion of the electroweak TUs in the $W$ mass measurement, 
including complete \Order{\alpha} electroweak (EW) corrections,
is not available  in the literature%
\footnote{For instance, it is missing in the CERN Report of the 2000
LEP2 MC Workshop~\cite{LEP2YR:2000}.}.
On the other hand, it is becoming a burning issue, as the error on the 
combined LEP2 result of the $W$ mass measurement approaches 30~MeV, while 
the total TU should be limited to $<15$~MeV.

Since the $W$ mass measurement has a very specific character, very
different from the measurement of the total cross section, 
let us characterize it briefly.
The actual way in which $M_W$ is measured by the LEP2 experiments is 
complicated, 
see e.g. refs.~\cite{Aleph-MW:2000,Delphi-MW:2001,L3-MW:1999,Opal-MW:2001}.
In particular, it seems to be beyond the reach of the simple fit 
to a one-dimensional $W$ invariant mass distribution%
\footnote{
  Even more inappropriate is trying to characterize the TU of the $W$ mass
  by introducing some kind of an ``error band'' in the one-dimensional 
  $W$ distribution -- see also the discussion below.}
(having integrated over the invariant mass of the second $W$).
This is due to direct inobservability of the neutrino (from the $W$ decay)
and of most of the initial-state radiation (ISR) photons,
loss of a fraction of the hadronic final state  in the beam pipe
and the non-trivial dependence on the invariant masses of both $W$'s
that certain corrections may have.

Let us give the reader at least a rough idea of how the
$W$ mass $M_W$ is measured by the LEP2 experiments.
In a nutshell, this is done with the help of a two-level fitting procedure.
At the first level, with the help of the so-called kinematic fit, an entire 
multi-momentum event, either experimental or of Monte Carlo (MC) origin,
is reduced to a point in much fewer dimensions than the total dimension
of the original set of four-momenta.
This space consists typically of the two fitted $W$ masses and of an auxiliary
parameter controlling the detector energy resolution.
In this way one gets 3-dimensional histograms with $\sim 10^4$
experimental events.
On the other hand, one gets the analogous 3-dimensional histograms
from a Monte Carlo simulation with $\sim 10^7$ events.
The latter one is obtained typically from the combined
KoralW~\cite{koralw:1995a,koralw:1998,koralw:2001} and  
YFSWW3~\cite{yfsww2:1996,yfsww3:1998,yfsww3:1998b,yfsww3:2000a,yfsww3:2001} 
programs, which we shall refer to as K-Y.
The actual ``$M_W$-extraction'' is done by fitting $M_W$
such that the difference between the above two 3-dimensional histograms
is minimized (typically using a likelihood function)%
\footnote{The above description of the ``$M_W$-extraction'' tries to summarize
  the methods used by ALEPH, L3 and OPAL; 
  the DELPHI method is slightly different, 
  see. ref.~\cite{Delphi-MW:2001}.}.
The K-Y prediction for every bin in the 3-dimensional histogram
is of course dependent on $M_W$.
This dependence is calculated/recalculated in the above fitting procedure
by means of averaging the ``correcting weight''~\cite{koralw:2001,yfsww3:2001}
corresponding to a variation of $M_W$,
over the entire $\sim 10^7$ sample of MC events stored on a computer disk. 
All complications of the experimental detector
and data analysis are therefore taken into account without any approximation.
In this way the multidifferential distribution implemented in the K-Y MC
ensures a direct unbiased link between the $M_W$
of the electroweak Lagrangian and the experimental LEP2 data,
assuming perfect detector simulation.

All this sounds like a strong argument to show that the
theoretical uncertainties, coming from higher-order
corrections and other imperfections of the theoretical calculations,
can be studied {\em only} within the programming environment used in the 
actual LEP2 experiments, with the help of the K-Y MC tandem.
Nevertheless, mainly because all effects under the following discussion are 
small, it makes sense to compromise and apply a
``simplistic approach'' based on the one-dimensional fit of a single $W$ 
effective mass (integrating over the second one).
This is what we shall do in the following.
The main danger in the use of a fitting procedure like the one described here
is that almost any physical effect in the $W$ effective mass distribution may 
feature strong correlations as a function of the two effective masses, 
which may lead to an underestimate of the effect by a factor of 2.
Our fitting method provides, therefore, a valuable but rough
estimate of the size of the effects under discussion in terms of 
the $M_W$ measured using LEP2 data.
{\em Consequently, if some effect turns out to be sizeable, that is
at least 1/3 of the experimental error on $M_W$ ($\sim 10$~MeV),
then it should be reanalysed within a full-scale ``$M_W$-extraction'' procedure
of the relevant LEP2 data analysis.} 
In such a case, our paper can be used as a guideline for a more complete
study to be  performed by the experiments.

Keeping all the above warnings and restrictions in mind, let us characterize
more precisely our aims and adopted methodology. 
During the 2000 LEP2 MC Workshop, the main emphasis was on the
TU for the total cross section ($\sigma_{tot}$)
of the $W$-pair production process~\cite{4f-LEP2YR:2000}.
Since a variation of $M_W$ is not related
to the overall normalization of the distribution 
$\rho(M_1,M_2)=d\sigma/(dM_1 dM_2)$ 
at $M_1=M_2=M_W$, but rather to the derivatives 
$D=(\partial/\partial M_i) \rho(M_1,M_2)|_{M_i=M_W}$,
the discussion of TUs on $M_W$ is almost completely 
independent of the discussion of TUs on $\sigma_{tot}$. 
{\em The higher order corrections, which strongly influence $\sigma_{tot}$,
may be completely unimportant for $M_W$ and vice versa!}
In particular, it is inappropriate to try to translate
our knowledge of TU in $\sigma_{tot}$, in terms of a certain ``error band''
in the distribution $d\sigma/dM_1$, into an error estimate of $M_W$ 
-- obviously it may easily lead to a {\em huge} overestimate of the TU of
$M_W$ and to overlooking effects which {\em really} contribute to it.

In the following, we consider the semileptonic process
$e^+ e^- \rightarrow u\bar{d} \mu^-\bar{\nu}_{\mu}$,
which belongs to the so-called CC11 class of Feynman diagrams
constituting the gauge-invariant subset of the 4-fermion final-state
processes, see e.g. ref.~\cite{LEP2YR:1996} for more details. 
We shall study only the leptonic $W$ mass, i.e. the one reconstructed 
from the four-momenta of the $\mu^-$ and $\bar{\nu}_{\mu}$
(in the actual experiments, the neutrino four-momentum is reconstructed
from the constrained kinematic fit, see e.g. 
refs.~\cite{Aleph-MW:2000,Delphi-MW:2001,L3-MW:1999,Opal-MW:2001}).
The input parameters are the same as in the 2000 LEP2 MC Workshop
studies~\cite{4f-LEP2YR:2000}. 
All the results in this paper are given for the centre-of-mass energy
$E_{CMS} = 200$~GeV and for the input $W$ mass $M_W = 80.350$~GeV.
The fitting function (FF) in all cases was taken from the semi-analytical 
program KorWan~\cite{koralw:1995a,koralw:1995b,koralw:1998}%
\footnote{The relevant distribution will be available in the next release of
  KorWan/KoralW.}.
All the results presented in the following, except the ones denoted with
the label {\tt 4f}, are for the leading-pole approximation (LPA)
of the Stuart-type~\cite{stuart:1997} (the LPA$_a$ option in YFSWW3) applied
to the above process, i.e. for the double-resonant $WW$ production and
decay (see ref.~\cite{yfsww3:2001} for more details). 

\begin{figure}[!ht]
\centering
\setlength{\unitlength}{0.1mm}
\epsfig{file=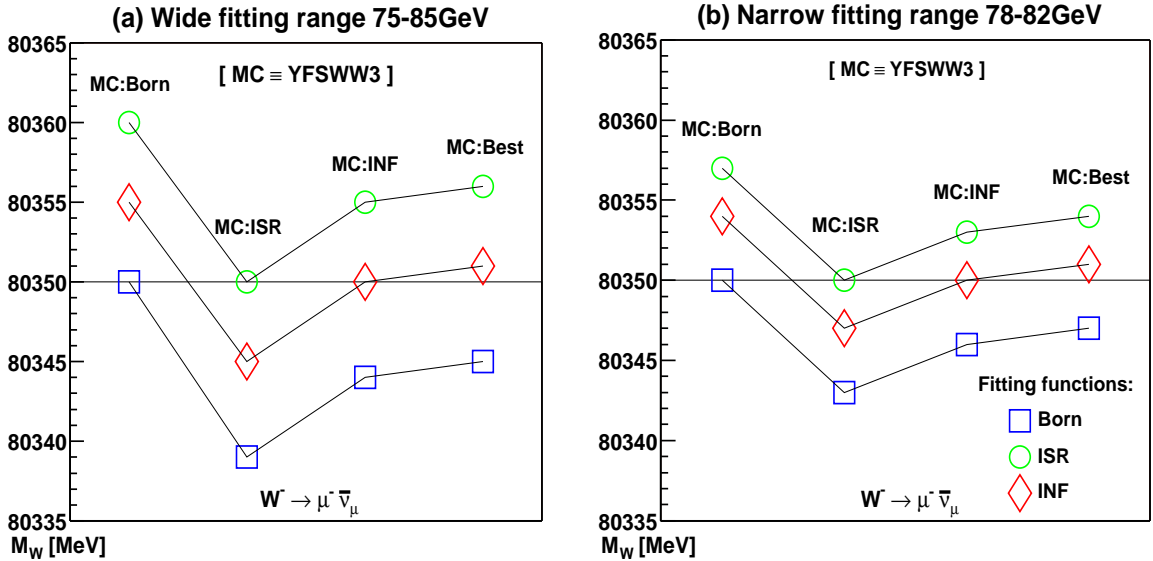,width=160mm,height=80mm}
\caption{\small\sf
  The introductory exercise, see more discussion in the text.
}
\label{fig:Fig1}
\end{figure}
In the first preparatory step, we construct a simple fitting procedure of the
$M_W$ using the 1-dimensional distribution of 
the $W$ effective mass $M_1$, and we ``calibrate'' with the help of 
the MC data in which we switch the same effects on and off, 
typically the ISR and the non-factorizable corrections (NF)
in the inclusive approximation (denoted by INF in the following) 
of the so-called screened Coulomb 
ansatz by Chapovsky \& Khoze~\cite{Chapovsky:1999kv},
just to see whether we get agreement in the case of the same effect 
in the MC data and in the fitting function. 
The other immediate profit is that we also quantify these effects as a shift of $M_W$.
The results of the first exercise are shown in Fig.~\ref{fig:Fig1}.
Let us explain briefly the notation:
{\tt Born} denotes the Born-level results, 
{\tt ISR} the ones including the \Order{\alpha^3} LL YFS 
exponentiation for the ISR 
as well as the standard Coulomb correction~\cite{khoze:1995}, 
{\tt INF} the above plus the INF correction, and 
{\tt Best} denotes the best predictions from YFSWW3, i.e. all the above 
plus the \Order{\alpha^1} electroweak non-leading (NL) corrections%
\footnote{The \Order{\alpha^1} electroweak corrections for the $WW$ production
          stage in YFSWW3 are based on 
          refs.~\cite{fleischer:1989,fleischer:1994}.}.

Let us summarize observations resulting from Fig.~\ref{fig:Fig1}:
\begin{itemize}
\item
  The fitted $M_W$ exactly agrees with the input $M_W$ in the case when the same 
  ISR and INF are included both in the fitting function (FF) and the MC.
\item
  If one is interested only in the shift of $M_W$, then any of the three FFs 
  could be used.
  In the following, in a single exercise we shall typically use one or two of 
  them only.
\item
  The dependence on the fitting range is sizeable; it points out, 
  albeit in a crude way, the fact discussed above, that the ultimate precise 
  fit of $M_W$ should always be done
  as in the LEP2 experiments, using a multidimensional fitting procedure.
\item
  The size of the ISR effect is about $-10$~MeV, that of the INF about 
  $+5$~MeV, 
  and the size of the NL corrections seems to be negligible, $\sim 1$~MeV.
\end{itemize}
Note that there were no cuts and we used the true parton-level 
$W$ invariant masses in all the above exercises.

In the following exercises we shall examine the influence of various 
effects/corrections on $M_W$ for various cuts and acceptances. 
Not all these effects can be included in the FF. 
Besides, only a very limited menu of cuts and acceptances can be applied
in the FF. Therefore,  our estimates of the TU will
based {\em not on absolute values} of the fitted $M_W$ but {\em on
relative differences} of $M_W$s corresponding to various effects.  
This is justified by our ``calibration'' exercise. 

\begin{figure}[!th]
\centering
\setlength{\unitlength}{0.1mm}
\epsfig{file=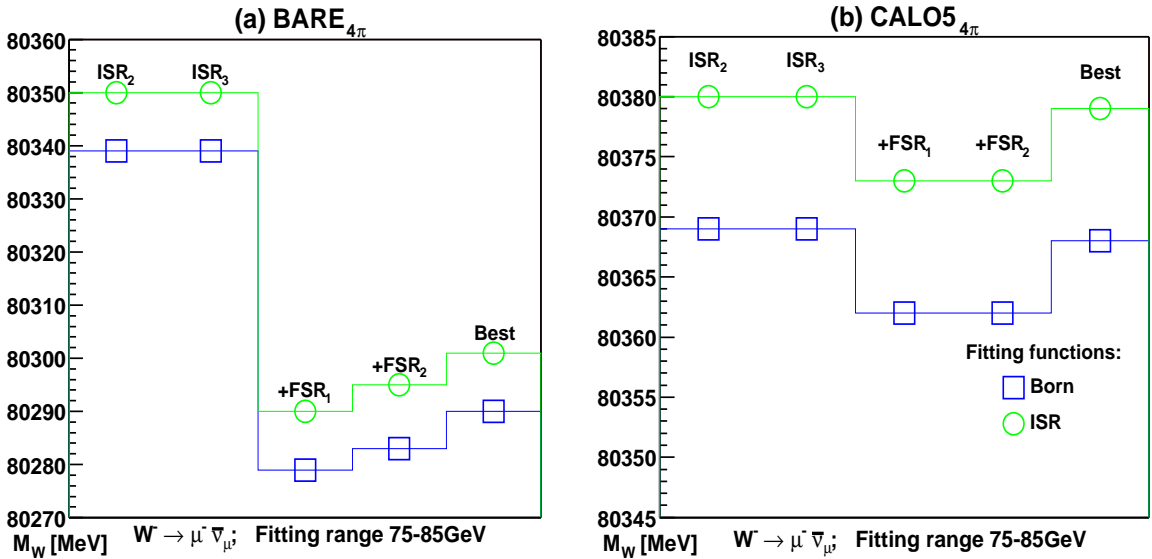,width=160mm,height=80mm}
\caption{\small\sf
  Effects of the ISR and the FSR on $M_W$.
}
\label{fig:Fig3}
\end{figure}

In the second exercise, depicted in Fig.~\ref{fig:Fig3}, 
we switch on and off various orders/variants of the ISR and of the
final-state radiation (FSR), and finally the NL correction.
The FSR was generated by the program PHOTOS~\cite{photos:1994}. 
While the previous exercise was without any cuts
and for the so-called BARE$_{4\pi}$ acceptance 
(the subscript $4\pi$ means the full solid-angle coverage),
we employ here the semi-realistic acceptance CALO5$_{4\pi}$,
where all the photons
for which the invariant mass with a final-state charged fermion was 
$<5$~GeV were recombined with that fermion 
(also for the full solid-angle coverage).
We compare the results for FF representing the Born and Born$+$ISR (no FSR)
levels.\\
Observations:
\begin{itemize}
\item
  Changing the type of the ISR from \Order{\alpha^3}$_{exp}$ LL
  to \Order{\alpha^2}$_{exp}$ LL induces a negligible, $<1$ MeV, effect 
  in fitted $M_W$.
\item
  The FSR effect is large for BARE$_{4\pi}$, $\sim 60$~MeV, and
  much smaller, $\sim 7$~MeV, for calorimetric CALO5$_{4\pi}$, as expected.
\item
  Switching from the single-photon (FSR$_1$) to the double-photon (FSR$_2$)
  option in PHOTOS results in a $\sim 4$~MeV change of $M_W$ for BARE$_{4\pi}$ 
  and no change for CALO5$_{4\pi}$.
\item
  The INF$+$NL correction is $\sim 6$~MeV and seems to cancel partly 
  with the FSR, see CALO5$_{4\pi}$.
\end{itemize}

Before we go to the next exercise, let us describe briefly the
acceptances and cuts
that were used in the MC simulations for the following calculations.
\begin{enumerate}
\item
  We required that the polar angle of any charged final-state fermion
  with respect to
  the beams be $\theta_{f_{ch}}>10^{\circ}$.
\item
  All photons within a cone of $5^{\circ}$ around the beams are treated
  as {\em invisible}, i.e. they were not included in the calculation of  
  the $W$ invariant masses.
\item
  The invariant mass of a {\em visible} photon with each 
  charged final-state fermion, $M_{f_{ch}}$, is calculated, 
  and the minimum value 
  $M^{min}_{f_{ch}}$ is found. If $M^{min}_{f_{ch}}< M_{rec}$ 
  or if the photon energy $E_{\gamma}<1\,$GeV,
  the photon is combined with the corresponding fermion, 
  i.e. the photon four-momentum is added to the fermion four-momentum
  and the photon is discarded. This is repeated for all {\em visible} 
  photons.\\
  In our numerical tests we used three values of the recombination cut: 
  \begin{displaymath}
  M_{rec} = \left\{
  \begin{tabular}{l}
   \: 0\,{\rm GeV:} \hspace{1cm} {\rm BARE}, \\
   \: 5\,{\rm GeV:} \hspace{1cm} {\rm CALO5}, \\
   25\,{\rm GeV:} \hspace{1cm} {\rm CALO25}.
  \end{tabular}  
  \right.
  \end{displaymath}
  Let us remark that we have changed here the labelling of these
  recombination cuts from the slightly misleading {\em bare} and {\em calo}
  names used in Ref.~\cite{4f-LEP2YR:2000}. They correspond to our
  CALO5 and CALO25, respectively.
  This change allows us to reserve the BARE
  name for a ``truly bare final fermion'' setup (without any recombination).
\end{enumerate}

\begin{figure}[!th]
\centering
\setlength{\unitlength}{0.1mm}
\epsfig{file=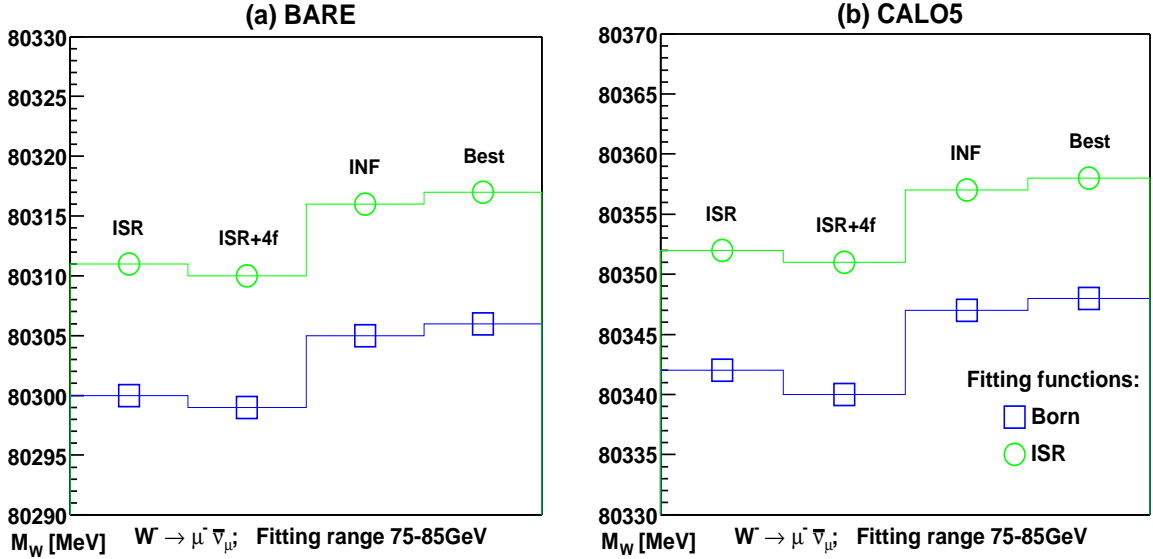,width=160mm,height=80mm}
\caption{\small\sf
  Effects of the $4f$-background, INF and NL corrections on $M_W$.
}
\label{fig:Fig4}
\end{figure}

In the next exercise, presented in  Fig.~\ref{fig:Fig4}, we examine once again
the effect of switching on the $4f$-background corrections%
  \footnote{The complete Born-level $4f$ matrix element
            in KoralW was generated with the help of the GRACE2 
            package~\cite{GRACE2}.}
and the INF corrections, now for BARE and CALO5.
The effect of the $4f$ background is $\sim 1$~MeV (it is therefore negligible 
for the LEP2 experiments%
\footnote{This smallness is due to a general smallness of the $4f$-background
 correction in the CC11 class of channels for LPA$_a$; it
 may be less pronounced for a different type of LPA, such as LPA$_b$
 in YFSWW3, for example.
 Our conclusion is unaffected, as it is really meant for the sum
 of LPA$_a$ and $4f$; see also the discussion below.})
and that of the INF is $\sim 5$~MeV.
The effects of the $4f$ background in the non-CC11 channels can be
larger, but they strongly depend on the applied experimental cuts or acceptances,
so that they can be studied in detail only within the full-scale LEP2 
$M_W$ fitting framework.

The size of the NF effect of $\sim 5$~MeV requires some explanation,
as the genuine NF effect in $M_W$ is in fact only about $\sim -1$~MeV;
see refs.~\cite{beenakker:1997,beenakker:1997b,dittmaier:1998,dittmaier:1998b}
and the discussion of the INF ansatz in ref.~\cite{Chapovsky:1999kv} 
(see also more discussion in the following).
This effect in Fig.~\ref{fig:Fig4}, understood as a difference between our ISR
and INF calculations,
is blown up artificially, because the ISR includes the $\sim -6$~MeV 
$M_W$ shift due to
the so-called ``standard Coulomb effect'' for historical reasons, 
although its derivation is not valid far away from the $WW$ threshold.

Another point to be explained is whether the genuine NF effect of  
$\sim -1$~MeV obtained in the INF (inclusive) approximation
can be increased to higher values, say $10$~MeV, due to the LEP2 experimental 
cuts. In principle it can be; however, as is well known,
the NF correction does not include (fermion) mass logarithms,
and its ``energy scale'', which enters the big logarithm owing to a cut on 
the photon energy, is $\Gamma_W$ and not $\sqrt{s}$.
Consequently, in order to get an enhancement factor of 
$\ln(\Gamma_W/E_{\max})\sim 10$,
one would need to veto the appearance of any photon above  $E_{\max}=0.1$~MeV 
-- a very unrealistic experimental selection indeed.
%
%
On the contrary, in the actual LEP2 experiments photons with energy 
$\le 2$~GeV are not disturbed, directly or indirectly, by the experimental 
event selection. This is why any strong enhancement of the NF effect
with respect to its ``inclusive'' treatment (INF) must be just absent.

\begin{figure}[!ht]
\centering
\setlength{\unitlength}{0.1mm}
\epsfig{file=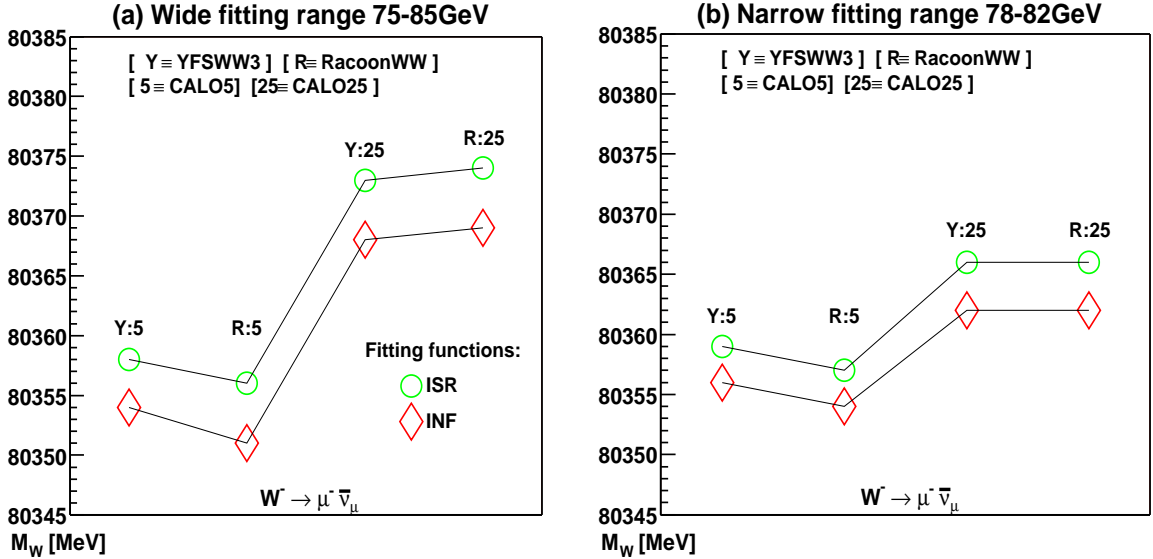,width=160mm,height=80mm}
\caption{\small\sf
  Comparison of YFSWW3 and RacoonWW.
}
\label{fig:Fig6}
\end{figure}
In the Fig.~\ref{fig:Fig6}, we examine the difference in $M_W$
fitted  to the $W$ mass distribution obtained from YFSWW3 and 
from RacoonWW~\cite{Denner:2000bj,Denner:2000bm}.
The distributions used for the $M_W$ fits
are exactly the same as those that were used for the plots
in the CERN Report of the 2000 LEP2 MC Workshop, see \cite{4f-LEP2YR:2000}. 
The statistical error is taken into account in the fits
and propagates into the fitted $W$ mass. It is merely $<1$~MeV.
We use two fitting functions, one in which the ISR is included
(with the incomplete NL but with the Coulomb effect)
and another one in which the INF (Chapovsky \& Khoze) is also included.
It is done for two kinds of calorimetric acceptances: the CALO5 and CALO25
described above.

Observations concerning the results shown in Fig.~\ref{fig:Fig6}:
\begin{itemize}
\item
  The comparison of YFSWW3 with RacoonWW is very interesting because
  the two calculations differ in almost every aspect of the implementation
  of the ISR, FSR, NL and NF corrections.
\item
  It is quite striking that
  the results of YFSWW3 and RacoonWW differ, in terms of the fitted mass, 
  by only $\le 3$~MeV, slightly more for CALO5 than for CALO25.
\item
  The difference between YFSWW3 and RacoonWW is definitely smaller 
  than the size  of the INF correction, roughly by a factor of 2
  (the INF is of order $3$--$5$~MeV for these two acceptances).
\end{itemize}
The most important result of the
comparison between YFSWW3 and RacoonWW is that it reconfirms the smallness of
the NF corrections in the $W$ mass.
Its size is  well below the $10$~MeV precision
target of the TU for the measurement of $M_W$ at LEP2.
It would be interesting to repeat the above exercise for the true LEP2
acceptance, using the full-scale fitting procedure.
In our opinion the above difference between YFSWW3 and RacoonWW in terms of 
the fitted $M_W$ cannot be attributed to some dominant source.
Most likely it consists of several contributions and the leading
candidates are ordinary {\em factorizable} QED corrections 
and/or some purely technical/numerical problems.

\begin{table}[!ht]
{
\begin{tabular}{||l|c|c|c||}
\hline\hline
\multicolumn{ 4}{|||c|||}{{ $\Delta M_W$ } }\\
\hline\hline
Error Type   & Scale Param. $\Delta M_W =\Gamma\times \epsilon$ 
& Numerical cross-check  &  $\Delta M_W$ \\
\hline
\multicolumn{ 4}{||c||}{{ $WW$ production } }\\
\hline
ISR ${\cal O}(\alpha^4 L_e^4)$
& $\epsilon \simeq \frac{\Gamma_W M_W}{s\beta_W^2} 
   (\frac{\alpha}{\pi})^4 L_e^4 
   \sim 5\cdot 10^{-6}$
& {\small 
  [${\cal O}(\alpha^3 L_e^3)-{\cal O}(\alpha^2 L_e^2)]_{\rm KoralW}$}
                                              & $\ll 1$~MeV \\
ISR ${\cal O}(\alpha^2 L_e)$
& $\epsilon \simeq \frac{\Gamma_W M_W}{s\beta_W^2} (\frac{\alpha}{\pi})^2 L_e 
   \sim 5\cdot 10^{-6} $
            & KorWan
                                              & $\ll 1$~MeV \\
ISR ${\cal O}(\alpha^2)_{pairs}$
& $\epsilon \simeq  \frac{\Gamma_W M_W}{s\beta_W^2} 
   (\frac{\alpha}{\pi})^2 L_e^2 
   \sim 4\cdot 10^{-4}$
            & KorWan
                                              & $< 1$~MeV \\
\hline
\multicolumn{ 4}{||c||}{{ $W$ decay } }\\
\hline
FSR ${\cal O}(\alpha)_{miss.}$
& $ \epsilon \simeq 0.2\,\Big(\frac{\pi}{8} \frac{\alpha}{\pi}
     2\ln\frac{M_W}{p_T}\Big)
     \sim 10^{-3}$
                   & Basic tests of PHOTOS
                                      & $\sim 2$~MeV\\
FSR ${\cal O}(\alpha^2)_{miss.}$
& $\epsilon \simeq \frac{1}{2}
  \Big(\frac{\pi}{8} \frac{\alpha}{\pi} 
  2\ln\frac{M_W}{p_T}\Big)^2 \sim 10^{-5}$
                   & On/off $2\gamma$ in PHOTOS
                                      & $\ll 1$~MeV \\
\hline
\multicolumn{4}{||c||}{ Non-factorizable QED interferences (between production
 and 2 decays)}\\
\hline
 ${\cal O}(\alpha^1)_{miss.}$ &
$\epsilon \simeq 0.1 \left(\frac{\alpha}{4}\frac{(1 - \beta)^2}{\beta}\right)
           \sim 10^{-4}$
                   & Chapovsky \& Khoze
                                     & $< 2$~MeV
\\
 ${\cal O}(\alpha^2)$ &
$\epsilon \simeq  \frac{1}{2}
 \left(\frac{\alpha^2}{4}\frac{(1 - \beta)^2}{\beta}\right)^2
            \sim  10^{-7}$
                   & None
                                     & $\ll 1$~MeV
\\
\hline\hline
\end{tabular}
}
\caption{\small\sf
  Estimation of the missing effects in the K-Y MC tandem.
}
\label{tab:Tab1}
\end{table}

In order to gain better understanding of the above numerical results it is
worth while to estimate them semi-quantitatively, in terms  of some 
``scale parameters'' representing various QED or EW corrections to $M_W$.
We shall do it in the following.
In addition, we shall discuss certain effects not included in K-Y or 
RacoonWW. The corrections to $M_W$ are generally of the type 
   $\frac{ \delta M_W}{M_W} \sim \frac{\Gamma_W}{M_W} \varepsilon$ 
or $\frac{ \delta M_W}{M_W} \sim \big(\frac{\Gamma_W}{M_W}\big)^2 \varepsilon$,
where $\varepsilon$ is a small parameter of the perturbative expansion.
We divide them into three types:
\begin{description}
\item[Case (a):]
  A mildly mass-dependent correction to the $W$ mass distribution
  $\rho(M)=\frac{d\sigma}{dM} \simeq |BW(M)|^2 \times f(M^2)$,
  which leads to 
  $\delta M_W \simeq \frac{1}{8} \Gamma_W^2  \frac{d\ln f(M^2)}{dM}|_{M=M_W}$,
  where $BW(M)$ denotes the Breit--Wigner resonance function and
  $f(M^2)$ is a mild function in the vicinity of the resonance
  (in the semi-quantitative discussion, we usually take $M=(M_1+M_2)/2$).
  The most trivial example is the kinematic factor 
  $f(M^2)=\beta_M = (1 - 4M^2/s)^{1/2}$, yielding 
  $\delta_{kin} M_W \simeq - \Gamma_W \frac{\Gamma_W M_W}{2s\beta_W^2}$,
  where $\beta_W = \beta_M|_{M=M_W}$.
  It is not visible in our fits (always taken into account in the FF); 
  however, the $\beta_W$-factor gets modified by the ISR, giving rise to
  $\delta M_W \simeq \delta_{kin} M_W \times 2\frac{\alpha}{\pi} L_e
  \simeq - 6$~MeV ($L_e = 2\ln\frac{s}{m_e^2}$)%
  \footnote{It is obtained from the approximate evaluation of the 
    derivative of the ISR convolution:
    $\frac{\partial \ln }{\partial M}\;
    \big[\int^1 dz \frac{ \beta_M(sz)}{\beta_M(s)} \gamma_{ISR} 
    (1-z)^{\gamma_{ISR}-1} \big]|_{M=M_W}$,
    where $\gamma_{ISR}= \frac{\alpha}{\pi} L_e$.
    }.
  This effect is responsible for most of the $M_W$ shift when switching 
  from the Born to the ISR in Fig.~\ref{fig:Fig1}. 
  It vanishes at high energies.
  The response of $\delta M_W$ to a more general variation:
  $f(x)\to f(x) +\varepsilon f_1(x)$, where $\varepsilon f_1$ is due
  to the higher-order ISR correction, is in general negligible, $< 1$~MeV.
  It can be estimated using
  $\delta M_W \simeq \varepsilon \frac{1}{8} \Gamma_W^2 
  \frac{d\ln f_1(M^2)}{dM}|_{M=M_W}
  \simeq \varepsilon \frac{\Gamma_W^2 M_W}{4s} $
  (here, we exploit the fact that $f_1$ has a derivative of \Order{1} as 
  a function of $M^2/s$).
  For instance, the missing ${\cal O}(\alpha^2)$ NLL ISR is proportional to 
  $\varepsilon \sim \alpha^2 L_e \simeq 10^{-3}$, giving rise to 
  $\delta M_W \sim 10^{-3}$~MeV.
  
\item[Case (b):]
  This is the case of the QED effects in the decays, the so-called FSR. 
  In this case the mass distribution gets distorted according to 
  $\frac{d\sigma}{dM^2}(M^2)
  \simeq \int dz\; \gamma_{FSR}(1-z)^{\gamma_{FSR}-1} 
  \frac{d\sigma}{dM^2}(z M^2)$,
  where $\gamma_{FSR}\simeq (\frac{\alpha}{\pi}) 
  \ln(M^2_W/m^2_{\mu}) \simeq 0.03$ for the BARE
  and $\gamma_{FSR}\simeq (\frac{\alpha}{\pi}) 
  \ln(M_W^2/m_{CALO}^2)\simeq 0.01$
  for CALO-type acceptance (with $m_{CALO}=5$~GeV).
  The mass shift $\delta M_W \simeq \Gamma_W \, \varepsilon$,
  $\varepsilon \simeq -\frac{\pi}{8}\,\gamma_{FSR} \simeq -0.012 $,
  is accounted for in the complete ${\cal O}(\alpha)$ calculation;
  see also refs.~\cite{Beenakker:1998cu,Beenakker:1998gr}.
  In the case of PHOTOS the missing ${\cal O}(\alpha)$ is related mainly 
  to high-$p_T$ photons,
  and from tests of this program listed in ref.~\cite{photos:1994}
  one can conclude that it corresponds to
  $\sim 0.2 \times \varepsilon,\; 
 \varepsilon \simeq  \frac{\alpha}{\pi} $. 
  The missing ${\cal O}(\alpha^2)$ FSR effect we estimate as follows:
  $\Delta \delta M_W \sim \Gamma_W \frac{1}{2}\varepsilon^2 \ll 1$ MeV.
  
\item[Case (c):] 
  The influence of the NF QED interferences on the $W$ mass is characterized
  by the correction function $\delta_{NF}$, which is strongly dependent 
  on $M_W$ in the vicinity of the resonance:
  $\frac{d\sigma}{dM} \simeq |BW(M)|^2 f(M) 
  \left[1 + \alpha\delta_{NF}\Big(\frac{M^2 - M_W^2}{M_W\Gamma_W}\Big)\right]$.
  The resulting $M_W$-shift is
  $\delta M_W \simeq \frac{1}{8}\Gamma_W^2 \alpha 
  \frac{d\delta_{NF}(M)}{dM}|_{M=M_W}$. 
  The simple INF formula for $\delta_{NF}$ of ref.~\cite{Chapovsky:1999kv}, 
  leads to
  $\delta M_W \simeq - \Gamma_W \frac{\alpha}{4}\frac{(1 - \beta_W)^2}{\beta_W}
  \sim -1$~MeV, in perfect agreement 
  with the complete calculations of the NF corrections, see 
  refs.~\cite{beenakker:1997,beenakker:1997b,dittmaier:1998,dittmaier:1998b}.
  If ${\cal O}(\alpha^1)$ NF is accounted for, then we estimate the missing 
  ${\cal O}(\alpha^2)$ NF contribution at
  $\Delta\delta M_W \sim \Gamma_W \frac{1}{2}\left(\frac{\alpha^2}{4}
  \frac{(1 - \beta_W)^2}{\beta_W}\right)^2 \ll 1$~MeV.
\end{description}
The above discussion confirms that all our numerical results are consistent
with expectations based on the ``scale parameters'' analysis and 
semi-quantitative calculations, and provides some estimates of the effects 
not accounted for in our MC programs.
It is also summarized in Table~\ref{tab:Tab1} together
with the relevant numerical estimates.

Let us finally discuss a question of the TU due to 
the so-called ``ambiguity of a definition of the LPA''.
In YFSWW3, we implemented two different definitions of the LPA, called
LPA$_a$ and LPA$_b$ \cite{yfsww3:2001}. 
Differences in $M_W$ reconstructed from the results obtained 
in these two options can give us a hint of the TU due to the LPA. 
Actually, we need to check only the variation of $M_W$ caused by the NL 
corrections. This is because KoralW implements the full $4f$-process 
at the so-called ISR level;
hence, the ambiguity  due the LPA is 
reduced from \Order{\frac{\Gamma_W}{M_W}} to 
\Order{\frac{\alpha}{\pi}\frac{\Gamma_W}{M_W}}
and is located only in the NL part.
We performed numerical tests of the dependence of NL on the choice of the LPA 
with the help of YFSWW3, finding the variation of the $M_W$
induced by the change from LPA$_a$ to LPA$_b$ to be $\leq 1$~MeV
(the ISR  and NL parts are always defined as in 
refs.~\cite{yfsww3:2001,koralw:2001}).
Another uncertainty in the LPA is due to the missing higher orders
in the NL part.
This can be estimated by comparing the predictions
of the so-called schemes {\em (A)} and {\em (B)}~\cite{yfsww3:2001} in YFSWW3.
These two schemes account for some higher-order effects by the 
use of the effective couplings in two different ways -- in fact, the scheme
{\em (B)} follows the prescription employed in RacoonWW.
We have checked that the change from the scheme {\em (A)} to the scheme
{\em (B)} results in the fitted $W$ mass shift of $\le 1$~MeV
(as expected, the results of the latter scheme are slightly closer to the ones
of RacoonWW).
Consequently, we attribute $\Delta M_W = 1$~MeV to the TU of $M_W$ due to 
the LPA.

In the above, we have considered only the leptonic $W$ mass coming from
the $W^- \rightarrow \mu^-\bar{\nu}_{\mu}$ decay. For the other
leptonic decays, the results should be similar when one applies the
calorimetric-type acceptance -- as was shown in ref.~\cite{yfsww3:1998b}.
For the hadronic $W$ masses at the parton level, the results would
be analogous to the ones presented here with the FSR switched off%
\footnote{Including the photon radiation from quarks without QCD effects
          is too crude an approximation and we do not consider such a scenario here.}.
More realistic estimates of the TU for the hadronic $W$ mass
would require taking into account the QCD effects, hadronization,
jet definitions, etc.
This should be done in the full-scale experimental data analysis,
which is beyond the scope of this paper.

From the above numerical exercises and the accompanying discussion, 
we come to the following conclusions:
\begin{itemize}
\item
  The electroweak theoretical uncertainty in $M_W$  
  of the K-Y MC tandem at LEP2 energies is $\sim 5$~MeV.
\item
  The above conclusion is strengthened by the smallness of the differences 
  between YFSWW3 and RacoonWW, which we attribute to the standard 
  {\em factorizable} corrections (ISR, FSR, etc.) 
  and purely technical/numerical effects.
\item
  In the above estimate we included a ``safety factor'' of $2$, corresponding 
  to the fact that our fits of $M_W$ were done for 1-dimensional 
  effective $W$ mass distributions.
  In order to eliminate it, our analysis should be repeated for 
  the realistic measurements of the LEP2 experiments.
\end{itemize}

\noindent
{\bf Acknowledgements}

We would like to thank R. Chierici, F. Cossutti and  A. Valassi
for useful discussions.
We also thank the authors of RacoonWW, A. Denner, S. Dittmaier, M. Roth 
and D. Wackeroth, for interesting discussions and for providing us their 
results.
We acknowledge the kind support of the CERN TH and EP Divisions.
One of us (B.F.L.W.) thanks Prof. S.~Bethke for the kind hospitality
and support of the Werner-Heisenberg-Institut, MPI, Munich,
and thanks Prof. C.~Prescott for the kind hospitality of SLAC Group A
while part of this work was done.

\bibliographystyle{prsty}


\end{document}